\documentclass[10pt,a4]{article}
\usepackage{mathrsfs,amsmath,amsthm,latexsym,amssymb,amsfonts,epsfig,dsfont,txfonts,undertilde}
\usepackage{hyperref}

\newcommand{\sst}{\scriptscriptstyle}

\def\d{\delta}
\def\sq{\sqrt{q}}
\def\f{\frac}
\def\ka{\kappa}
\def\ep{\epsilon}
\def\ue{\utilde{\epsilon}}
\def\te{\tilde{\epsilon}\,}
\def\mx{{\mathrm{d}}^3x}
\def\my{{\mathrm{d}}^3y}
\def\ms{{\mathrm{d}}s}
\def\mt{{\mathrm{d}}t}
\def\t{\tilde}
\def\lp{\ell_{\rm p}}
\def\rd{{\rm d}}
\def\sgn{{\rm sgn}}
\allowdisplaybreaks
\begin{document}

\title{{\sf New length operator for loop quantum gravity}}

\author{
Yongge Ma$^1$\thanks{mayg@bnu.edu.cn},~~ Chopin
Soo$^2$\thanks{cpsoo@mail.ncku.edu.tw} ~~ and ~ Jinsong
Yang$^2$\thanks{yangksong@gmail.com}\vspace{0.5em}\\
$^1$Department of Physics, Beijing Normal University, Beijing
100875,
P. R. China\\
$^2$Department of Physics, National Cheng Kung University, Tainan
70101, Taiwan \\}
\date{}

\maketitle

\begin{abstract}
{\sf An alternative expression for the length operator in loop
quantum gravity is presented. The operator is background
independent, symmetric, positive semidefinite, and well defined on
the kinematical Hilbert space. The expression for the regularized
length operator can moreover be understood both from a simple
geometrical perspective as the average of a formula relating the
length to area, volume and flux operators, and also consistently as
the result of direct substitution of the densitized triad operator
with the functional derivative operator into the regularized
expression of the length. Both these derivations are discussed, and
the origin of an undetermined overall factor in each case is also
elucidated. }
\end{abstract}

{\hspace{1em}PACS numbers: 04.60.Pp, 04.60.Ds}

\section{Introduction}

Geometrical quantities such as the length, area and volume are of
great importance to general relativity (GR) as a theory of
geometrodynamics. In loop quantum gravity (LQG) (for reviews on the
subject, see for instance, Refs.
\cite{ThomasRev,RovRev,ALRev,HanRev}) these operators have been
realized as well-defined quantum operators on the kinematical
Hilbert space, and they have been demonstrated to have discrete
spectra in Refs.
\cite{ThomasL,BianchiL,R-S-Volum,A-L-Area,A-L-Volum,ThomasV}. The
Bekenstein-Hawking entropy of a black hole, for instance, was
computed in LQG based on the quantum area operator and the concept
of isolated horizons, and the volume operator was used in
constructing mathematically well-defined Hamiltonians which
determine the quantum dynamics in Ref. \cite{QSDI}. There are also
well-defined energy operators, the Arnowitt-Deser-Misner energy
operator in Ref. \cite{QSDVI} and the quasilocal energy operators in
Refs. \cite{Y-M-Energy,Mar-Energy}. Moreover, in Ref.
\cite{Y-M-Energy} the Geroch energy operator was used to derive a
rather general entropy-area relation and thus a holographic
principle from loop quantum gravity.

In this work a new expression for length operator in LQG will be
presented. The expression for the operator can be understood as
originating from a simple geometrical formula relating the length to
area, volume, and flux operators. In Euclidean 3D space, the length
of an edge in a hexahedron can be expressed as a composition of the
area of 2-surface, volume of region and angle between 2 surfaces in
the hexahedron (see, for instance, Fig. \ref{partition}). While
Refs. \cite{ThomasL,BianchiL} have defined length operators in the
framework of loop quantum gravity, a simple geometrical relationship
between the length and the area, volume and angles is not
articulated.

We first present a regularized length operator by transcribing the
classical relation into a formula that inherits this simple
geometrical composition which can be expressed directly in terms of
the fundamental elements---area, volume, fluxes---of LQG. Our first
method of regularization is along the lines that led to the volume
operator in Ref. \cite{A-L-Volum}. The final expression for the
background-independent length operator contains an undetermined
overall factor which arises from the process of averaging. This is
also encountered in the construction of the volume operator in Ref.
\cite{A-L-Volum}. We use ``internal" regularization (similar to the
regularization of volume in Ref. \cite{A-L-Volum}, and so called
because the regulated identity (in Sec. \ref{subsec}) is expressed
in terms of densitized triads smeared over two surfaces within the
interior of the cell). This is different from the ``external"
regularization introduced in Ref. \cite{BianchiL} (the length
operator therein is, without averaging, dependent on background
structures). In Ref. \cite{ThomasL} a length operator was
constructed through a different strategy---substituting $e^i_{a}
\propto \{A^i_a, V\}$ in the classical length identity before
regularization and quantization.

We demonstrate in a second derivation that the length operator can
also be obtained by direct substitution of the densitized triad
operator (similar to the method in Ref. \cite{ThomasV}) as the
functional derivative with respect to the connection into the
regularized expression. An undetermined overall factor is shown to
arise from the choice of a characteristic function employed in the
derivation and regularization.

Although the above two routes to the length operator are rather
distinct, up to the overall factor, the final expression for the
background-independent quantum length operator is reassuringly
identical.

\section{Elements of LQG}\label{lqg}
We briefly recap the elements of LQG to establish our notations and
conventions. The Hamiltonian formalism of GR is formulated on a
4-dimensional manifold $M=\mathbb{R}\times\Sigma$, with $\Sigma$
being a 3-dimensional manifold of arbitrary topology. Introducing
Ashtekar-Barbero variables \cite{AshVar,Barbero}, GR can be cast as
a dynamical theory with $SU(2)$ ($SO(3)$ for pure GR) connection.
The phase space consists of canonical pairs $(A^i_a, \t{E}^a_i)$ of
fields on $\Sigma$, where $A^i_a$ is a connection 1-form which takes
values in the Lie algebra $su(2)$, and $\t{E}^a_i$ is a vector
density of weight 1. Spatial indices are denoted by $a,b,c,...$ and
$i,j,k,...=1,2,3$ are internal indices. The densitized triad
$\t{E}^a_i$ is related to the cotriad $e_a^i$ by
$\t{E}^a_i=\frac{1}{2}\,\te^{abc}\ep_{ijk}e^j_be^k_c\sgn(\det(e^i_a))$,
wherein $\te^{abc}$ is the Levi-Civit\`a tensor density of weight 1,
and $\sgn(\det(e^i_a))$ denotes the sign of $\det(e^i_a)$. The
3-metric on $\Sigma$ is expressed in terms of cotriads through
$q_{ab}=e^i_ae^j_b\delta_{ij}$. The only nontrivial Poisson bracket
is given by
\begin{align}
\{A^i_a(x),\t{E}^b_j(y)\}=\ka\beta\d^b_a\d^i_j\d^3(x,y),
\end{align}
with $\ka=8\pi G$, and $\beta$ is the Barbero-Immirzi parameter.

The fundamental variables in LQG are the holonomy of the connection
along a curve and the flux of densitized triad through a 2-surface.
Given a curve $c:[0,1]\rightarrow\Sigma$, the holonomy $h_c(A)$ of
connection $A^i_a$ along the curve $c$ is
\begin{align}
 h_c(A)={\cal P}{\rm
 exp}\left(\int_cA\right)=1_2+\sum_{n=1}^{\infty}\int_0^1\mt_1
 \cdots\int_{t_1}^1\mt_2\cdots\int_{t_{n-1}}^1\mt_nA(c(t_1))\cdots
 A(c(t_{n})),
\end{align}
wherein ${\cal P}$ denotes the path ordering which orders the
smallest path parameter to the left,
$A(c(t)):=\dot{c}^a(t)A^j_a(c(t))\tau_j/2$,  $\dot{c}^a(t)$ is the
tangent vector of $c$, and $\tau_j=-i\sigma_j$ (with $\sigma_j$
being the Pauli matrices). The flux $\t{E}_j(S)$ of densitized triad
$\t{E}^a_j$ through a 2-surface $S$ is explicitly
\begin{align}
  \t{E}_j(S)=\int_Sn^S_a\t{E}^a_j,
\end{align}
with $n^S_a$ being the conormal vector with respect to the surface
$S$.

Another element of LQG is the notion of edges and graphs embedded in
$\Sigma$ (see, for instance, \cite{ThomasRev} for a review). An edge
$e$ is an equivalence class of curves $c_e$ which is semianalytic in
all of [0, 1] . By $\gamma$ we denote a closed, piecewise analytic
graph which is a set of edges that intersect at most in their end
points. The collection of all end points of edges in a graph
$\gamma$ is denoted by $V(\gamma)$, while the set of all edges in
$\gamma$ is denoted $E(\gamma)$. In order to simplify the notation,
we subdivide each edge $e$ with endpoints $v,v'$ which are vertices
of $\gamma$ into two segments $e_1,e_2$ wherein $e=e_1\circ
(e_2)^{-1}$, with $e_1$ having an orientation that it is {\em
outgoing} at $v$ and the orientation of $e_2$ is also outgoing at
$v'$. This introduces new vertices $e_1\cap e_2$ which we will call
pseudovertices because they are not points of nonanalyticity of the
graph. We still denote the set of these segments of $\gamma$ by
$E(\gamma)$ for simplicity, but the set of true (as opposed to
pseudo) vertices of $\gamma$ will be denoted as $V(\gamma)$.\

To construct quantum kinematics, one has to extend the configuration
space ${\cal A}$ of smooth connections to the space $\bar{\cal A}$
of distributional connections. A function $f$ on $\bar{\cal A}$ is
said to be cylindrical with respect to a graph $\gamma$ iff it can
be written as $f=f_\gamma\circ p_{\gamma}$, wherein
$p_\gamma(A)=(h_{e_1}(A),..,h_{e_n}(A))$ and $e_1,..,e_n$ are the
edges of $\gamma$. Here $h_e(A)$ is the holonomy along $e$ evaluated
at $A\in\bar{\cal A}$ and $f_\gamma$ is a complex-valued function on
$SU(2)^n$. Since a function cylindrical with respect to a graph
$\gamma$ is automatically cylindrical with respect to any graph
bigger than $\gamma$, a cylindrical function is actually given by a
whole equivalence class of functions $f_\gamma$. We will henceforth
not distinguish between this equivalence class and one of its
representatives in the set of cylindrical functions denoted by ${\rm
Cyl}(\bar{\cal A})$.

Through projective techniques, $\bar{\cal A}$ is equipped with a
natural, faithful, ``induced" measure $\mu_0$, called the
Ashtekar-Isham-Lewandowski measure \cite{AI,AL}. In a certain sense,
this measure is the unique diffeomorphism-invariant measure on
$\bar{\cal A}$ \cite{Uniq}, and the kinematical Hilbert space is
then ${\cal H}_{\mathrm{kin}}=L^2(\bar{\cal A},{\rm{d}}\mu_0)$. The
cylindrical function space ${\rm Cyl}(\bar{\cal A})$ is a dense
subset of ${\cal H}_{\mathrm{kin}}=L^2(\bar{\cal A},{\rm{d}}\mu_0)$,
and cylindrical functions act by multiplication and fluxes by
derivation on ${\cal H}_{\mathrm{kin}}=L^2(\bar{\cal
A},{\rm{d}}\mu_0)$. Given a graph $\gamma$ and a 2-surface $S$, we
can change the orientations of some edges of $\gamma$ and subdivide
edges of $\gamma$ into two halves at an interior point if necessary,
and obtain a graph $\gamma_{\sst S}$ adapted to $S$ such that the
edges of $\gamma_{\sst S}$ belong to the following four types
\cite{ThomasRev}: (i) $e$ is up with respect to $S$ if $e\cap
S=e(0)$ and $\dot{e}^a(0)n^S_a(e(0))>0$; (ii) $e$ is down with
respect to $S$ if $e\cap S=e(0)$ and $\dot{e}^a(0)n^S_a(e(0))<0$;
(iii) $e$ is inside with respect to $S$ if $e\cap S=e$; (iv) $e$ is
outside with respect to $S$ if $e\cap S=\emptyset$. The flux
operator $\hat{\t{E}}_j(S)$ acting on a function $f$ cylindrical
with respect to a graph $\gamma$ adapted to $S$ is given by
\begin{align}
  \hat{\t{E}}_j(S)\cdot f=-\f{i\beta\lp^2}{4}\sum_{e\in
  E(\gamma)}\varrho(e,S)X^j_e\cdot f_\gamma,
\end{align}
wherein $\lp^2=\kappa\hbar$, $X^j_e={\rm
tr}\left(\tau_jh_e(A)\f{\partial}{\partial h_e(A)}\right)$ is the
right invariant vector field, and $\varrho(e,S)$ takes values of
$0$, $+1$ and $-1$ corresponding to whether the edge $e$ is
inside/outside, up or down with respect to the surface $S$.

\section{The length operator}

In this section, a new length operator for LQG will be presented. To
wit, let ${\cal C}$ be the set of continuous, oriented, piecewise
semianalytic, parametrized, compactly supported curves embedded into
$\Sigma$, wherein a curve $c\in {\cal C}$ can be parametrized by
\begin{align}
c:[0,1]\rightarrow \Sigma;\quad s\mapsto c(s).
\end{align}
Then the length of the curve $c$ is given by
\begin{align}\label{defleng}
  L(c)&=\int_0^1\ms\sqrt{q_{ab}(c(s))\dot{c}^a(s)\dot{c}^b(s)}=\int_0^1\ms
  \sqrt{e^i_a(c(s))e^j_b(c(s))\d_{ij}\dot{c}^a(s)\dot{c}^b(s)}
  \equiv\int_0^1\ms\sqrt{\d_{ij}l^i(s)l^j(s)}\,,
\end{align}
wherein $q_{ab}$ is the metric of $\Sigma$, and $l^i(s)\equiv
e^i_a(c(s))\dot{c}^a(s)$.

In  what follows, we will regularize and quantize the length
expression using two different methods following the strategies in
\cite{A-L-Volum} and \cite{ThomasV}. The first method can be easily
visualized geometrically, and the second has the advantage of being
more direct and compact in its derivation.

\subsection{The first strategy}
We first define and quantize the length of a curve by adapting the
framework in \cite{A-L-Volum}. Our task is to regularize the
expression of length in \eqref{defleng} into an expression which is
suitable for quantization.

\subsubsection{Regularization procedure for length identity}\label{subsec}
\begin{figure}[htbp]
\centering
\begin{minipage}[b]{0.32\textwidth}
\begin{tabular}{c}
\includegraphics[height=1.15\textwidth]{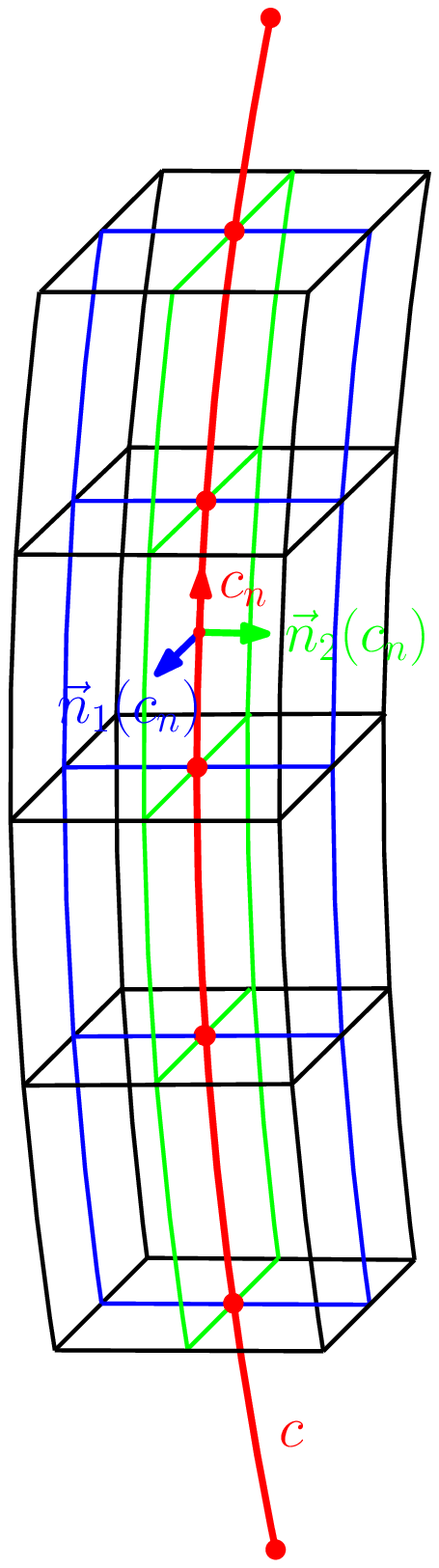}\\
(a)
\end{tabular}
\end{minipage}
\begin{minipage}[b]{0.32\textwidth}
\begin{tabular}{c}
\includegraphics[height=1.15\textwidth]{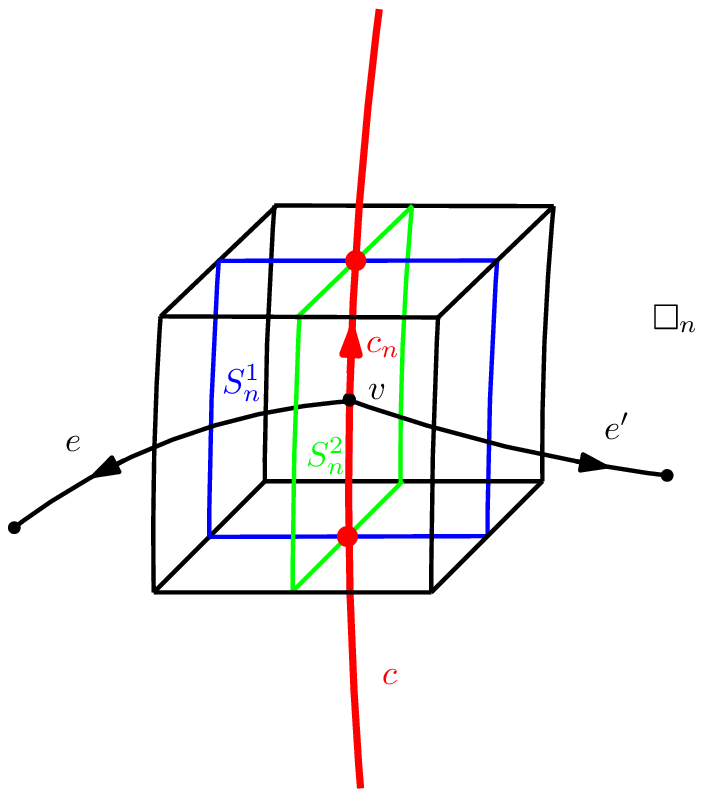}\\
(b)
\end{tabular}
\end{minipage}
\caption{(a) A partition of the neighborhood $R_c$ of the curve $c$
in $\Sigma$ into cubes. (b) The figure illustrates the way two
2-surfaces of a partition are adapted to the curve $c_n$ according
to conditions (i) and (ii) with the vertex $v$ at the center.}
\label{partition}
\end{figure}

The regularization procedure involves the following ingredients.
Partitioning of the curve $c$ as a composition of $N$ segments
$\{c_n\}, n\in\mathds{N}, 1\leq n\leq N$, i.e.,
\begin{align}
  c=c_1\circ c_2\circ\cdots\circ c_n\circ\cdots\circ c_N,
\end{align}
wherein $\circ$ is a composition of composable curves which can be
carried out with
\begin{align}
  c_n:[(n-1)\ep,n\ep]\rightarrow \Sigma;\quad s_n\mapsto c_n(s_n),
\end{align}
and $\ep=\f1N$. The second ingredient involves a partition of the
neighborhood $R_c$ of the curve $c$ in $\Sigma$. To do that, let us
first fix global coordinates $x^a=(\sigma^1,\sigma^2,s)$ for $R_c$,
and for each segment $c_n$ introduce two surfaces $S_n^I,I=1,2$
intersecting in $c_n$ defined by
$S^1_n=S^1_n(\sigma^1=0,\sigma^2,s_n)$ and
$S^2_n=S^2_n(\sigma^1,\sigma^2=0,s_n)$ (with orientation induced by
that of the coordinate axes). Let $n_a^I(c_n):=(\rd\sigma^I)_a,
I=1,2$, denote the dual normal vector field of $S^I_n$. For each
$c_n$, a close cube $\Box_n$ containing $c_n$ (determined by $S^1_n$
and $S^2_n$) gives then a partition of $R_{c_n}$ adapted to $c_n$
(see Fig. \ref{partition} (a)).
For fixed $\ep$, we assume that the coordinate areas $\ep L^I_n$ of
$S^I_n$ are bounded from above by $\ep^2$ (i.e., $L^I_n<\ep$).
Notice that
\begin{align}
  l^i(s_n)&=\left.\f12\ue_{abc}\ep^{ijk}\t{E}^b_j\t{E}^c_k\dot{c}_n^a/\sq\,\right|_{c_n(s_n)}
   =\left.\f14\ue_{abc}\ep^{ijk}\t{E}^b_j\t{E}^c_k\ep_{IJ}\te^{ade}n^I_dn_e^J/\sq\,\right|_{c_n(s_n)}
     =\left.\f12\,\ep_{IJ}\ep^{ijk}n^I_b\t{E}^b_jn_c^J\t{E}^c_k/\sq\,\right|_{c_n(s_n)},
\end{align}
wherein the identity
\begin{align}
  \dot{c}_n^a(s_n)=\left.\f12\,\ep_{IJ}\te^{abc}n^I_bn_c^J\right|_{c_n(s_n)}.
\end{align}
has been made use of, $\utilde{\epsilon}_{abc}$ and $\epsilon_{IJ}$
are the 3D and 2D Levi-Civit\`a tensor densities of weight $-1$,
respectively. We can define the length segment via
\begin{align} \label{lf}
  L^\ep(c_n)&:=\sqrt{\d_{ij}l^i_{n,\ep}l^j_{n,\ep}}\,,
\quad\text{with}\quad
  l^i_{n,\ep}=\f{\f12\,\ep_{IJ}\ep^{ijk}\t{E}_j(S^I_n)\t{E}_k(S^J_n)}{V(\Box_n)},
\end{align}
and $V(\Box_n)$ is the volume of $\Box_n$. The expression
$L^\ep(c_n)$ yields the length $L(c)$ in \eqref{defleng} of the
curve $c$ as
\begin{align}\label{Lc}
  L(c)&=\lim_{\ep\rightarrow0}\sum_{n=1}^NL^\ep(c_n)
  =\lim_{\ep\rightarrow0}\sum_{n=1}^N\sqrt{\f{\ep_{IJ}\ep_{KL}\t{E}_j(S^I)\t{E}_k(S^J)\t{E}_j(S^K)\t{E}_k(S^L)}{2\left[V(\Box_n)\right]^2}}\notag\\
  &=\lim_{\ep\rightarrow0}\sum_{n=1}^N\sqrt{\f{\t{E}_j(S^1_n)\t{E}_j(S^1_n)\t{E}_k(S^2_n)\t{E}_k(S^2_n)-\left[\t{E}_j(S^1_n)\t{E}_j(S^2_n)\right]^2}
   {\left[V(\Box_n)\right]^2}}\notag\\
   &=\lim_{\ep\rightarrow0}\sum_{n=1}^N\sqrt{\f{\left[Ar(S_n^1)Ar(S_n^2)\right]^2-\left[\t{E}_j(S^1_n)\t{E}_j(S^2_n)\right]^2}
   {\left[V(\Box_n)\right]^2}}\,,
\end{align}
wherein $Ar(S_n^I)$ is precisely the area of the surface $S^I_n$.
The regulated length $L^\ep(c_n)$ depends on the classical phase
space variables through the flux, the areas of 2-surfaces and the
volume of 3D region which have direct correspondence to quantum
operators in LQG. Hence it is straightforward to promote
$L^\ep(c_n)$ to its quantum version $\hat{L}^\ep(c_n)$.

A geometrical picture for the regularized expression of length in
Eq. \eqref{Lc} can be given. Let us first recall the angle operator
introduced in \cite{Major-Direc}. At $c_n$, the angle
$\theta_{(c_n(s_n),S^1_n,S^2_n)}$ between $n_a^1(c_n(s_n))$ and
$n_a^2(c_n(s_n))$ measured with respect to the 3D metric $q_{ab}$ is
given by
\begin{align}\label{angle}
  \cos\theta_{(c_n(s_n),S^1_n,S^2_n)}&=\f{q^{ab}n_a^1n_b^2}{\sqrt{q^{cd}n_c^1n_d^1}\sqrt{q^{ef}n_e^2n_f^2}}\,(c(s_n))
  =\lim_{S^1_n,S^2_n\rightarrow
  c_n(s_n)}\f{\t{E}_j(S^1_n)\t{E}_j(S^2_n)}{Ar(S^1_n)Ar(S^2_n)}.
\end{align}
With Eq. \eqref{angle}, the expression in Eq. \eqref{Lc} reduces to
\begin{align}\label{relation}
  L(c)&=\lim_{\ep\rightarrow0}\sum_{n=1}^N\sqrt{\f{\left[Ar(S_n^1)Ar(S_n^2)\right]^2
       -\left[Ar(S_n^1)Ar(S_n^2)\cos\theta_{(c_n(s_n),S_n^1,S_n^2)}\right]^2}{\left[V(\Box_n)\right]^2}}\notag\\
       &=\lim_{\ep\rightarrow0}\sum_{n=1}^N\f{Ar(S_n^1)Ar(S_n^2)\left|\sin\theta_{(c_n(s_n),S_n^1,S_n^2)}\right|}{V(\Box_n)}.
\end{align}
In our regularization procedure for the length operator, the virtue
of the expression of length in \eqref{relation} (or \eqref{Lc}) is
that it is closely related to other geometric operators (area,
volume, and angle) which have well-defined quantum actions.
Moreover, for a rectangular hexahedron $\Box_n$ in flat space,
$L^\ep(c_n)$ is just the familiar expression of the length of the
vertical edge $c_n$ of $\Box_n$.

The above regularization of the length is so-called internal because
the regulated identity \eqref{lf} is expressed in terms of triads
smeared over two surfaces passing the interior of the cell and the
regularization matches the internal regularization of volume in
\cite{A-L-Volum}. An external regularization of the length of a
curve was investigated in \cite{BianchiL}.

\subsubsection{Quantum length operator}
The volume operator in LQG has been thoroughly discussed in Refs.
\cite{R-S-Volum,A-L-Volum,ThomasV}, and in this work we shall focus
on the volume operator in \cite{A-L-Volum} (\cite{ThomasV}).
Properties of the volume operator have been investigated in
\cite{Brunn}. The action of the volume operator measuring a region
$R$ on a function $f$ cylindrical with respect to $\gamma$ is given
specifically by
\begin{align}
  \hat{V}(R)\cdot f&=\sum_{v\in V(\gamma)\cap R}\hat{V}_v\cdot f_{\gamma}=\ell_{\rm
  p}^3\sum_{v\in V(\gamma)\cap R}\sqrt{\left|i Z
      \sum_{e_I\cap e_J\cap e_K=v}\varsigma(e_I,e_J,e_K)\ep_{ijk}X^i_{e_I}X^j_{e_J}X^k_{e_K}\right|}\,\cdot
      f_{\gamma}\,,
\end{align}
wherein $\varsigma(e_I,e_J,e_K)={\rm
sgn}(\det(\dot{e}_I(0),\dot{e}_J(0),\dot{e}_K(0)))$, and $Z$ the
ambiguity factor due to regularization has been uniquely fixed as
$1/(3!\times8\times8)$ in \cite{G-T-check} through consistency
checks of volume and triad operator regularizations. In the
regulated length $L^\ep(c_n)$ in \eqref{Lc} we shall need to define
the inverse volume operator. The volume operator can have a large
kernel, so the naive inverse volume operator $1/\hat{V}$ is not well
defined, but following \cite{BianchiL}, an inverse volume operator
based on the idea in \cite{Tik} may be taken to be
\begin{align}\label{invV}
  \widehat{V^{-1}}:=\lim_{\ep\rightarrow0}(\hat{V}^2+\ep^2\ell_{\rm
  p}^6)^{-1}\hat{V}\,.
\end{align}
This inverse volume operator has the same properties as the volume
operator in that in the limit
$\lim_{\ep\rightarrow0}\widehat{V^{-1}}(\Box_n)$ acts only on the
vertices of $\gamma$ of cylindrical function $f$ which are on $c_n$.
Let us consider the partition of $R_c$ adapted to a graph. More
precisely, we will assume that for sufficiently small $\epsilon$ the
permissible partitions satisfy the following conditions: (i)
$\Box_n$ contains at most one vertex of $\gamma$ which lies on $c_n$
as the interior point; (ii) if $v$ (lying on $c_n$) is the vertex of
$\gamma$, it is the unique isolated intersection point between the
union of the two 2-surfaces $S^I_n$ associated to $\Box_n$ and
$\gamma$ (see Fig. \ref{partition} (b)). Thus
$\widehat{V^{-1}}(\Box_n)$ vanishes when $\gamma$ has no vertex
contained in $\Box_n$. Including the nontrivial case of a vertex $v$
of $\gamma$ lying on $c_n$ the action of the quantum operator
$\hat{l}_{n,\ep}^i$ (corresponding to $l_{n,\ep}^i$ in \eqref{lf})
is then
\begin{align}
  \hat{l}^i_{n,\ep}\cdot
  f&=\widehat{V(\Box_n)^{-1}}\f12\ep_{IJ}\ep^{ijk}\hat{\t{E}}_j(S^I_n)\hat{\t{E}}_k(S^J_n)\cdot
  f\notag\\
  &=-\f{\beta^2\lp^4}{32}\chi_{\sst\Box_n}(v)\widehat{V^{-1}}_v\ep_{IJ}\ep^{ijk}\sum_{e(0)=v}\varrho(e,S^I_n)X_e^j\sum_{e'(0)=v}\varrho(e',S^J_n)X_{e'}^k
  \cdot f_\gamma\notag\\
  &=-\f{\beta^2\lp^4}{32}\chi_{\sst\Box_n}(v)\widehat{V^{-1}}_v\sum_{e(0)=e'(0)=v}\ep_{IJ}\varrho(e,S^I_n)\varrho(e',S^J_n)\ep^{ijk}X_e^jX_{e'}^k\cdot f_\gamma\notag\\
  &\equiv \chi_{\sst\Box_n}(v) \hat{l}^i_v\cdot f,
\end{align}
wherein $\chi_{\sst\Box_n}(v)$ is the characteristic function which
takes the value 1 when $v$ is contained in $\Box_n$ (and is zero
otherwise). Also $\hat{L}^\ep(c_n)$ (corresponding to $L^\ep(c_n)$
in \eqref{lf}) can now be defined as
\begin{align}
\hat{L}^\ep(c_n)\cdot
f&:=\sqrt{\hat{l}^i_{n,\ep}\left(\hat{l}^i_{n,\ep}\right)^\dag}\,\cdot
f=\chi_{\sst\Box_n}(v)\sqrt{\hat{l}^i_v\left(\hat{l}^i_v\right)^\dag}\,\cdot
f.
\end{align}
The length operator $\hat{L}(c)$ is then
\begin{align}
  \hat{L}(c)=\lim_{\ep\rightarrow0}\sum_n\hat{L}^\ep(c_n).
\end{align}

A choice has been made in the ordering of the noncommuting operators
in the expression $\hat{l}^i_v$ wherein the volume operator has been
ordered to the left, so alternatives up to these ordering
ambiguities are perhaps also viable.

The action of $\hat{L}^\ep(c_n)$ depends on the two 2-surfaces
$S^I_n$ only through the properties of these surfaces at $v$. Hence,
it is unchanged as we refine the partition and shrink the cell
$\Box_n$ to $v$ and the limit is thus
\begin{align}\label{qlc}
  \hat{L}(c)\cdot f=\sum_{v\in V(\gamma)\cap
  c}\sqrt{\hat{l}^i_v\left(\hat{l}^i_v\right)^\dag}\,\cdot
f_\gamma.
\end{align}
However, Eq. \eqref{qlc} carries information of our choice of
partitions through the terms
$\ep_{IJ}\varrho(e,S^I_n)\varrho(e',S^J_n)$ which depend on the
background structure---the coordinates choice defines the surface
$S^I_n$---just as was encountered in constructing the volume
operator in Ref. \cite{A-L-Volum}. Hence, although the limit of
$\hat{L}^\ep(c_n)$ is well defined, it is not yet viable as a
background-independent length operator. We can remove the background
structure by suitably ``averaging" the regularized operator over it
following the strategy in \cite{A-L-Volum}, and obtain the average
of $\ep_{IJ}\varrho(e,S^I_n)\varrho(e',S^J_n)$ as $k_{\rm
av}\varsigma(c,e,e')$ wherein $k_{\rm av}$ is a constant, and
$\varsigma(c,e,e')$ is the orientation function which equals $+1$
$(-1)$ if the tangential directions of $c,e,e'$ are linearly
independent at the vertex $v$ and oriented positively( or
negatively), or zero otherwise. The averaging yields a final
well-defined background-independent LQG length operator $\hat{L}(c)$
\begin{align}\label{qlcav}
  \hat{L}(c)\cdot f=\sum_{v\in V(\gamma)\cap
  c}\sqrt{\hat{l}^i_{v,\rm{av}}\left(\hat{l}^i_{v,\rm{av}}\right)^\dag}\,\cdot f_\gamma,\qquad\text{with}\quad
\hat{l}^i_{v,\rm{av}}=-\f{k_{\rm
av}\beta^2\lp^4}{32}\,\widehat{V^{-1}}_v\sum_{e(0)=e'(0)=v}\varsigma(c,e,e')\ep^{ijk}X^j_eX^k_{e'}.
\end{align}

\subsection{The second strategy}

An alternative method is to derive the length operator for LQG by
adapting the technique developed in \cite{ThomasV}. The advantage of
this approach is its directness. Given a curve $c$, we may choose a
coordinate system $\{x^1,x^2,x^3\}$ for the neighborhood in
$\Sigma$, and let $\chi_{\sst\Delta}(p,x)$ be the characteristic
function in the coordinate $x$ of a cube with center $p$ spanned by
three vectors ${\pmb{\Delta}}_i=\Delta_i{\pmb{n}}_i$, wherein
${\pmb{n}}_i$ is the normal vector in the frame under consideration.
The cube has coordinate volume ${\rm
vol}(\Delta)=\Delta_1\Delta_2\Delta_3
\det({\pmb{n}}_1,{\pmb{n}}_2,{\pmb{n}}_3)$, wherein
$\det({\pmb{n}}_1,{\pmb{n}}_2,{\pmb{n}}_3)$ denotes the determinant
of the three normal vectors in the chosen coordinates (we assume the
three normal vectors to be right oriented). Explicitly,
\begin{align}
\chi_{\sst\Delta}(p,x)=
\prod_{i=1}^3\theta(\f{\Delta_i}{2}-|<n_i,x-p>|)\notag
\end{align}
with $<.,.>$ being the standard Euclidean inner product, and
$\theta(y)=1$ for $y>0$ and zero otherwise. In what follows we can
arbitrarily and smoothly extend the curve $c:[0,1]\rightarrow\Sigma;
s\mapsto c(s)$ to $\t{c}:[0-\lambda,1
+\lambda]\rightarrow\Sigma;t\mapsto\t{c}(t)$ (here $\lambda>0$) such
that $\t{c}(t)=c(s)$ for $t=s\in[0,1]$, and denote $\t{c}$ by $c$ in
the computations for simplicity. With
$\chi_\ep(s,t)=\theta(\f{\ep}{2}-|s-t|)$ wherein $s\in[0,1]$, we can
then construct the smeared quantity
\begin{align}\label{eD}
  l^i(s,\ep,\Delta,\Delta')&=\int_{-\lambda}^{1+\lambda}\mt\int_\Sigma
  \mx\int_\Sigma\my\f{\chi_\ep(s,t)}{\ep}\;\f{\chi_{\sst\Delta}\left(2c(s),c(t)+x\right)}{{\rm vol}(\Delta)}
  \f{\chi_{\sst\Delta'}\left(3c(s),c(t)+x+y\right)}{{\rm vol}(\Delta')}\notag\\
  &\qquad\times\f{1}{2\sqrt{\det(q)(x)}}\,\utilde{\ep}_{abc}\ep^{ijk}E^b_j(x)E^c_k(y)\,\dot{c}^a(t)\,.\notag\\
  &\approx\f{1}{2{\rm
  vol}(\Delta')}\int_{-\lambda}^{1+\lambda}\mt\int_\Sigma
  \mx\int_\Sigma\my\,\f{\chi_\ep(s,t)}{\ep}\chi_{\sst\Delta}\left(2c(s),c(t)+x\right)\chi_{\sst\Delta'}\left(3c(s),c(t)+x+y\right)\notag\\
  &\qquad\times\f{1}{V(x,\Delta)}\,\utilde{\ep}_{abc}\ep^{ijk}\dot{c}^a(t)E^b_j(x)E^c_k(y),
\end{align}
wherein $V(x,\Delta)$ denotes the volume of the cube
$\Box^{\sst\Delta}_x$ with center point $x$ and coordinate volume
${\rm vol}(\Delta)$, measured with respect to $q_{ab}$. The smeared
quantity $l^i(s,\ep,\Delta,\Delta')$ in \eqref{eD} reduces to
$l^i(s)$ in \eqref{defleng} when one takes the limit $\ep, \Delta,
\Delta'\rightarrow0$. Hence the length of curve $c$ in
\eqref{defleng} can be written as
\begin{align}
  L(c)&=\lim_{\ep,\ep'\rightarrow0}\lim_{\Delta,\Delta',\Delta'',\Delta'''\rightarrow0}\int_0^1\ms
   \sqrt{\d_{ij}l^i(s,\ep,\Delta,\Delta')l^j(s,\ep',\Delta'',\Delta''')}\;.
\end{align}
Again the problem of the naive inverse of the volume can be
circumvented by adopting
\begin{align}\label{inversV}
  \widehat{V(x,\Delta)^{-1}}\cdot f&:=\lim_{\ep'\rightarrow0}\left(\f{1}{\hat{V}(x,\Delta)^2+\ep'^2\ell^6_{\rm
  p}}\right)\hat{V}(x,\Delta)\cdot f_\gamma
  =\lim_{\ep'\rightarrow0}\sum_{v\in V(\gamma)\cap \Box^{\sst\Delta}_x}\left(\f{1}{\hat{V}^2_v+\ep'^2\ell^6_{\rm
  p}}\right)\hat{V}_v\cdot f_\gamma
  =:\sum_{v\in V(\gamma)\cap
  \Box^{\sst\Delta}_x}\widehat{V^{-1}}_v\cdot f_\gamma.
\end{align}
To proceed from \eqref{eD} to the quantum formula, we require just
two steps: the first is to promote the classical inverse volume to
its quantum version in \eqref{inversV}, and the second is to replace
$E^a_i$ by $\hat{\t{E}}^a_i(x)=-i\beta\ell_{\rm p}^2\d/\d A^i_a(x)$.
The regularized expression \eqref{eD} involves the quantum operator
\begin{align}
  \hat{\t{E}}^a_i(x,\Delta):=\f{1}{{\rm
vol}(\Delta)}\int_\Sigma\my\,\chi_{\sst\Delta}(x,y)\hat{\t{E}}^a_i(y),
\end{align}
which has a convenient action on a cylindrical function $f$. With
respect to the graph $\gamma$, the result is
\begin{align}\label{be}
\hat{\t{E}}^a_i(x,\Delta)\cdot f&=-\f{i\beta\lp^2}{2{\rm
vol}(\Delta)}\sum_{e\in
E(\gamma)}\int_{[0,1]}\mt\,\chi_{\sst\Delta}(x,e(t))\dot{e}^a(t){\rm
tr}\left(h_e(0,t)\tau_ih_e(t,1)\f{\partial}{\partial
h_e(0,1)}\right)\cdot f_\gamma.
\end{align}
wherein $h_e(t,t'), t<t'$ denotes the holonomy of $A^i_a$ along the
segment $[t,t']\rightarrow\Sigma$ of $e$. We can next evaluate the
action of $\hat{l}^i(s,\ep,\Delta,\Delta')$ on $f$. There are two
types of terms in the final result: the first type comes from only
the action of two functional derivatives on $f$, while the second
comes from that of one functional derivative acting on $f$ and the
other acting on the trace in \eqref{be}. Explicitly we have
\begin{align}\label{he}
\hat{l}^i(s,\ep,\Delta,\Delta')\cdot f
 &=-\f{\beta^2\lp^4}{8{\rm
  vol}(\Delta')}\int_{-\lambda}^{1+\lambda}\mt\int_{[0,1]^2}\mt'\mt''\left\{\sum_{e',e''\in
 E(\gamma)}\ue_{abc}\,\dot{c}^a(t)\dot{e}'^b(t')\dot{e}''^c(t'')\right.\notag\\
 &\quad\qquad\times\f{\chi_\ep(s,t)}{\ep}\chi_{\sst\Delta}\left(2c(s),c(t)+e'(t')\right)\chi_{\sst\Delta'}\left(3c(s),c(t)+e'(t')+e''(t'')\right)
 \widehat{V(e'(t'),\Delta)^{-1}}\notag\\
 &\quad\qquad\times\ep^{ijk}\mathrm{tr}\left(h_{e'}(0,t')\tau_jh_{e'}(t',1)\f{\partial}{\partial h_{e'}(0,1)}\right)
 \mathrm{tr}\left(h_{e''}(0,t'')\tau_jh_{e''}(t'',1)\f{\partial}{\partial
 h_{e''}(0,1)}\right)\notag\\
 &\qquad{\pmb+}\sum_{e'\in
 E(\gamma)}\utilde{\ep}_{abc}\,\dot{c}^a(t)\dot{e}'^b(t')\dot{e}'^c(t'')\ep^{ijk}\notag\\
 &\quad\qquad\times\chi_\ep(c(s),c(t))\chi_{\sst\Delta}\left(2c(s),c(t)+e'(t')\right)\chi_{\sst\Delta'}\left(3c(s),c(t)+e'(t')+e'(t'')\right)
 \widehat{V(e'(t'),\Delta)^{-1}}\notag\\
 &\quad\qquad\times\left[\theta(t''-t')\mathrm{tr}\left(h_{e'}(0,t')\tau_jh_{e'}(t',t'')\tau_kh_{e'}(t'',1)\f{\partial}{\partial
   h_{e'}(0,1)}\right)\right.\notag\\
  &\quad\qquad\quad\left.+\left.\theta(t'-t'')\mathrm{tr}\left(h_{e'}(0,t'')\tau_jh_{e'}(t'',t')\tau_kh_{e'}(t',1)\f{\partial}{\partial
  h_{e'}(0,1)}\right)\right]\right\}\cdot f_\gamma\notag\\
 &\equiv-\f{\beta^2\lp^4}{8{\rm
  vol}(\Delta')}\int_{-\lambda}^{1+\lambda}\mt\int_{[0,1]^2}\mt'\mt''\notag\\
 &\quad\times\left\{\sum_{e',e''\in
 E(\gamma)}\utilde{\ep}_{abc}\,\dot{c}^a(t)\dot{e}'^b(t')\dot{e}''^c(t'')\widehat{V(e'(t'),\Delta)^{-1}}\hat{O}^i_{ce'e''}(t,t',t'')\right.\notag\\
 &\qquad\qquad\qquad\qquad\times\f{\chi_\ep(s,t)}{\ep}\chi_{\sst\Delta}\left(2c(s),c(t)+e'(t')\right)\chi_{\sst\Delta'}\left(3c(s),c(t)+e'(t')+e''(t'')\right)\notag\\
 &\qquad\quad{\pmb+}\sum_{e'\in
 E(\gamma)}\utilde{\ep}_{abc}\,\dot{c}^a(t)\dot{e}'^b(t')\dot{e}'^c(t'')\widehat{V(e'(t'),\Delta)^{-1}}\hat{O'}^i_{ce'e'}(t,t',t'')\notag\\
 &\qquad\qquad\qquad\qquad\left.\times\f{\chi_\ep(s,t)}{\ep}\chi_{\sst\Delta}\left(2c(s),c(t)+e'(t')\right)\chi_{\sst\Delta'}
  \left(3c(s),c(t)+e'(t')+e'(t'')\right)\right\}\cdot f_\gamma.
\end{align}
Given a triple $(c,e',e'')$ consisting of a curve $c$ and two edges
$e',e''$ in $E(\gamma)$ (which contains the case $e'=e''$), we may
consider the vector valued function
\begin{align}
  z_{c,e',e''}(t,t',t''):=c(t)+e'(t')+e''(t'').
\end{align}
Its Jacobian
\begin{align}
  \det\left(\f{\partial\left(z^1_{ce'e''},z^2_{ce'e''},z^3_{ce'e''}\right)(t,t',t'')}{\partial(t,t',t')}\right)
  =\,\utilde{\ep}_{abc}\,\dot{c}^a(t)\dot{e}'^b(t')\dot{e}''^c(t'')
\end{align}
is precisely the factor that appears in all the integrals in
\eqref{he}. This is also the motivation for introducing the special
characteristic function
$\chi_{\sst\Delta'}\left(3c(s),c(t)+x+y\right)$ in \eqref{eD}.

Let us first take the limit $\Delta'\rightarrow0$. The integrand in
\eqref{he} vanishes unless $c(s)$ is a vertex $v=e'\cap e''$ of the
graph $\gamma$ in the limit. On the other hand, in order to make the
determinant nonvanishing at $c(s)$, the two edges $e',e''$ and the
curve $c$ must be distinct from one another. Hence, the second term
in \eqref{he}, which involves only summation over $e'(=e'')$,
vanishes. Now by letting $\Delta'$ be sufficiently small, the
condition $\chi_{\sst\Delta'}\left(c(s),z_{ce'e''}\right)=1$ implies
$\chi_{\sst\Delta}\left(2c(s),c(t)+e'(t')\right)=\chi_\ep(s,t)=1$,
so that we can pull the remaining characteristic functions
$\chi_{\sst\Delta}\left(2c(s),c(t)+e'(t')\right)$ and
$\chi_\ep(s,t)$ out of the integral by replacing them with
$\chi_{\sst\Delta}(c(s),v)=1$ and $\chi_\ep(s,c^{-1}(v))$,
respectively. For the case $c\cap e'\cap e''=c(s)=v$ is an outgoing
point (i.e. the parameters of the edges $e'$ and $e''$ in our
notation described in Sec. \ref{lqg} take the value 0), we can also
replace the operators $\widehat{V(e'(t'),\Delta)^{-1}}$ and
$\hat{O}^i_{ce'e''}(t,t',t'')$ by $\widehat{V(v,\Delta)^{-1}}$ and
$\hat{O}^i_{ce'e''}(c^{-1}(v),0,0)$. Hence \eqref{he} reduces to
\begin{align}
  \lim_{\Delta'\rightarrow0}\hat{l}^i(s,\ep,\Delta,\Delta')\cdot f
  &=-\f{\beta^2\lp^4}{8}\sum_{v\in V(\gamma)\cap c}\f{\chi_\ep(s,c^{-1}(v))}{\ep}\,\widehat{V(v,\Delta)^{-1}}\notag\\
  &\qquad\qquad\times\sum_{e'\cap e''=v}\;\hat{O}_{ce'e''}(c^{-1}(v),0,0)
  \int_{-\lambda}^{1+\lambda}\mt\int_{[0,1]^2}\mt'\mt''\d^3(c(s),x)\det\left(\f{\partial(z^a_{ce'e''})}{\partial(t,t',t'')}\right)\cdot f_\gamma\notag\\
  &=-\f{\beta^2\lp^4}{8}\sum_{v\in V(\gamma)\cap c}\f{\chi_\ep(s,c^{-1}(v))}{\ep}\,\widehat{V(v,\Delta)^{-1}}
  \sum_{e'\cap e''=v}\;\hat{O}^i_{ce'e''}(c^{-1}(v),0,0)\varsigma(c,e',e'')
  \int {\rm d}^3z\d^3(c(s),z)\cdot f_\gamma\notag\\
  &=-\f{\beta^2\lp^4}{8\times4}\sum_{v\in V(\gamma)\cap c}\f{\chi_\ep(s,c^{-1}(v))}{\ep}\,\widehat{V(v,\Delta)^{-1}}
    \sum_{e'\cap
    e''=v}\;\varsigma(c,e',e'')\hat{O}^i_{ce'e''}(c^{-1}(v),0,0)\cdot f_\gamma,
\end{align}
wherein
$\varsigma(c,e',e''):=\sgn(\det(\dot{c}(c^{-1}(v)),\dot{e}'(0),\dot{e}''(0))$,
$\hat{O}^i_{ce'e''}(c^{-1}(v),0,0):=\ep^{ijk}X^j_{e'}X^k_{e''}$, and
the factor $1/4$ came from the fact that the integral just equals
$\int_{-\lambda}^{1+\lambda}\mt\int_{[0,1]^2}\mt'\mt''\d(s,t)\d(0,t')\d(0,t'')=1/4$.
Using
$\lim_{\Delta\rightarrow0}\widehat{V(v,\Delta)^{-1}}=\widehat{V^{-1}}_v$
from \eqref{inversV}, we can easily take the $\Delta\rightarrow0$
limit to yield
\begin{align}
  \lim_{\Delta\rightarrow0}\lim_{\Delta'\rightarrow0}\hat{l}^i(s,\ep,\Delta,\Delta')\cdot
  f
  &=-\f{\beta^2\lp^4}{32}\sum_{v\in V(\gamma)\cap c}\f{\chi_\ep(s,c^{-1}(v))}{\ep}\,\widehat{V^{-1}}_v\sum_{e'\cap e''=v}
   \;\varsigma(c,e',e'')\hat{O}^i_{ce'e''}(c^{-1}(v),0,0)\cdot f_\gamma\notag\\
  &\equiv\sum_{v\in V(\gamma)\cap
  c}\f{\chi_\ep(s,c^{-1}(v))}{\ep}\;\hat{l}^i_v\cdot f_\gamma.
\end{align}
It follows that the regularized length operator can be defined as
\begin{align}
\hat{L}^{\ep,\ep'}(c)\cdot
f&:=\lim_{\Delta,\Delta',\Delta'',\Delta'''\rightarrow0}\int_0^1\ms
\sqrt{\hat{l}^i(s,\ep,\Delta,\Delta')\left(\hat{l}^i(s,\ep',\Delta'',\Delta''')\right)^\dag}\cdot f\notag\\
&=\int_0^1\ms\sqrt{\sum_{v\in V(\gamma)\cap
c}\f{\chi_\ep(s,c^{-1}(v))}{\ep}\,\hat{l}^i_v\sum_{v'\in
V(\gamma)\cap
c}\f{\chi_{\ep'}(s,c^{-1}(v'))}{\ep'}\left(\hat{l}^i_{v'}\right)^\dag}\;\cdot
f_\gamma.
\end{align}
For small enough $\ep,\ep'$ and a given $s$,
$\chi_\ep(s,c^{-1}(v))\chi_{\ep'}(s,c^{-1}(v'))$ vanishes unless
$v=v'$. Choosing $\ep=\ep'$, and for sufficiently small $\ep$, the
regularized length operator simplifies to
\begin{align}
\hat{L}^\ep(c)_\gamma=\int_0^1\ms\sqrt{\sum_{v\in V(\gamma)\cap
c}\left(\f{\chi_\ep(s,c^{-1}(v))}{\ep}\right)^2\hat{l}^i_v(\hat{l}^i_v)^\dag}.
\end{align}
Noting that for sufficiently small $\ep$ and any given $s$ there is
at most one vertex contribution i.e. $\chi_\ep(s,c^{-1}(v))\neq0$
for at most one vertex $v$, we can equivalently evaluate the sum as
\begin{align}
\hat{L}^\ep(c)_\gamma=\int_0^1\ms\sum_{v\in V(\gamma)\cap
c}\f{\chi_\ep(s,c^{-1}(v))}{\ep}\sqrt{\hat{l}^i_v(\hat{l}^i_v)^\dag}.
\end{align}
Finally, we can remove the regulator, i.e. take the limit as
$\ep\rightarrow 0$, and obtain our LQG length operator as
\begin{align}\label{2volumeop}
\hat{L}(c)_\gamma=\sum_{v\in V(\gamma)\cap
c}\sqrt{\hat{l}_v^i(\hat{l}_v^i)^\dag}, \qquad\text{with}\quad
  \hat{l}^i_v:=-\f{\beta^2\lp^4}{32}\,\widehat{V^{-1}}_v\sum_{e'\cap
  e''=v}\varsigma(c,e',e'')\ep^{ijk}X^j_{e'}X^k_{e''}.
\end{align}
Expression \eqref{2volumeop} is well defined, as it does not depend
on any background information. However, there is a freedom of choice
in the characteristic function (a similar situation also occurs in
the construction of the volume operator in \cite{ThomasV}). We may
replace $\chi_{\Delta'}(3c(s),c(t)+x+y)$ by
$\chi_{\Delta'}(3c(s),a_1c(t)+a_2x+a_3y)$ where $a_{i=1,2,3}$ are
arbitrary nonvanishing positive real numbers satisfying $\sum_i a_i
=3$ (so that the integrand in (28) in the limit $\Delta' \rightarrow
0$ vanishes unless $c(s)$ is a vertex $v = e'(0)\cap e''(0)$) in the
regularized expression $l^i(s,\ep,\Delta,\Delta')$ in \eqref{eD}.
This results in the final quantum length operator being
\begin{align}\label{lengop2}
\hat{L}(c)_\gamma=\sum_{v\in V(\gamma)\cap c}\sqrt{\hat{l}_{v,{\rm
alt}}^i(\hat{l}_{v,{\rm alt}}^i)^\dag}, \qquad\text{with}\quad
  \hat{l}^i_{v,{\rm alt}}:=-\f{k_{\rm alt}\beta^2\lp^4}{32}\,\widehat{V^{-1}}_v\sum_{e'\cap
  e''=v}\varsigma(c,e',e'')\ep^{ijk}X^j_{e'}X^k_{e''}.
\end{align}
wherein $k_{\rm alt}\equiv1/(a_1a_2a_3)$. Comparing \eqref{lengop2}
to \eqref{qlcav}, we are happy to observe that our two strategies of
regularization lead to essentially the same length operator.

\section{Concluding remarks}

So far we have constructed a family of operators
$(\hat{L}(c)_\gamma,D_\gamma)_{\gamma\in \Gamma}$, where
$D_\gamma={\rm Cyl}^3(\overline{\cal A})$ denotes the domain of
$\hat{L}(c)_\gamma$; but the proof for the cylindrical consistency
(this yields a length operator $\hat{L}(c)$ on ${\cal
H}_{\mathrm{kin}}$) of the family of operators is the same as that
for the volume operator (see, for instance, Ref. \cite{ThomasV}), so
we shall omit it here. Regularization ambiguities arise rather
frequently in the quantization procedure. Especially, different
regularization strategies would lead to different operators.
However, in our length operator two quite different regularization
procedures lead to essentially the same final expression. This
strengthens our confidence in the construction.

Although the length operator is well defined and background
independent, there exists an overall undetermined factor $k_{\rm
av}$ (or $k_{\rm alt}$) arising from averaging over the relevant
structure in the first derivation (or through the choice of the
characteristic function employed in the second derivation) which
needs to be fixed. In order to fix the ambiguity, two strategies can
be adopted. The first strategy is through the semiclassical limit of
the length operator - for instance, the recent semiclassical
analysis of the volume operator developed in \cite{F-semiVol} can
also be applied to our length operator. The second method is through
a consistency check. Since $\hat{l}_{v,{\rm av}}^i$ in \eqref{qlcav}
(or $\hat{l}_{v,{\rm alt}}^i$ in \eqref{lengop2}) is just an
alternative expression of the triad operator defined on a vertex
$v$,  the consistency check of volume and triad operator
quantizations (as was carried out in \cite{G-T-check} to fix the
ambiguity factor of the volume operator in \cite{A-L-Area} or
\cite{ThomasV}) provides a useful way to fix the ambiguity factor.

The classical length expression in \eqref{defleng} is invariant
under gauge transformations of the SU(2) group. This symmetry is
preserved in the quantum theory, whence, our length operator is
internal gauge invariant and hence can be defined in the internal
gauge invariant Hilbert space. For smooth diffeomorphisms on
$\Sigma$, the length operators transform covariantly.

In the expression of the length operator,
$\hat{l}_v^i(\hat{l}_v^i)^\dag$ is symmetric and positive
semidefinite on ${\cal H}_{\mathrm{kin}}$, and thus the square root
exists. Hence all $\hat{L}(c)_\gamma$ in the family
$(\hat{L}(c)_\gamma,D_\gamma)_{\gamma\in \Gamma}$ are positive
semidefinite, so the projective limit $\hat{L}(c)$ is a densely
defined, positive semidefinite and symmetric operator which has
self-adjoint extensions (for instance, its Friedrich extension).

\section*{Acknowledgments}

Y. M. is supported in part by NSFC No. 10975017; C. S. and J. Y. are
supported in part by the National Science Council of Taiwan under
Grant Nos. NSC 98-2112-M-006-006-MY3 and 98-2811-M-006-035, and by
the National Center for Theoretical Sciences, Taiwan.


\begin{thebibliography}{99}
\bibitem{ThomasRev}
T.\ Thiemann, Modern canonical quantum general relativity, Cambridge
University Press, Cambridge, 2007

\bibitem{RovRev}
C.\ Rovelli, Quantum gravity, Cambridge University Press, Cambridge,
2004

\bibitem{ALRev}
A.\ Ashtekar and J.\ Lewandowski, Background independent quantum
gravity: A status report,
\href{http://dx.doi.org/10.1088/0264-9381/21/15/R01}{Class.\ Quantum
Grav. {\bf 21} (2004) R53}

\bibitem{HanRev}
M.\ Han, W.\ Huang and Y. Ma, Fundamental structure of loop quantum
gravity, \href{http://dx.doi.org/10.1142/S0218271807010894}{Int.\
J.\ Mod.\ Phys. D {\bf 16} (2007) 1397}

\bibitem{ThomasL} T. Thiemann, A length operator for canonical
quantum gravity, \href{http://dx.doi.org/10.1063/1.532445}{J.\
Math.\ Phys. {\bf 39} (1998) 3372}

\bibitem{BianchiL} E.\ Bianchi, The length operator in Loop Quantum Gravity,
\href{http://dx.doi.org/10.1016/j.nuclphysb.2008.08.013}{Nucl.\ Phys.\ B {\bf 807} (2009)
591}

\bibitem{R-S-Volum}
C.\ Rovelli and L.\ Smolin, Discreteness of area and volume in
quantum gravity,
\href{http://dx.doi.org/10.1016/0550-3213(95)00150-Q}{Nucl.\ Phys.\
B {\bf 442} (1995) 593}

\bibitem{A-L-Area}
A.\ Ashtekar and J.\ Lewandowski, Quantum theory of geometry I: Area
operators,
\href{http://dx.doi.org/10.1088/0264-9381/14/1A/006}{Class.\ Quantum
Grav. {\bf 14} (1997) A55}

\bibitem{A-L-Volum}
A.\ Ashtekar and J.\ Lewandowski, Quantum theory of geometry II:
Volume operators, Adv.\ Theor.\ Math.\ Phys. {\bf 1} (1998) 388

\bibitem{ThomasV} T.\ Thiemann, Closed formula for the matrix
elements of the volume operator in canonical quantum gravity,
\href{http://dx.doi.org/10.1063/1.532259}{J.\ Math.\ Phys. {\bf 39}
(1998) 3347}

\bibitem{QSDI}
T.\ Thiemann, Quantum spin dynamics (QSD),
\href{http://dx.doi.org/10.1088/0264-9381/15/4/011}{Class.\ Quantum
Grav. {\bf 15} (1998) 839}

\bibitem{QSDVI}
T.\ Thiemann, Quantum spin dynamics (QSD): VI. Quantum Poincar\'e
algebra and a quantum positivity of energy theorem for canonical
quantum gravity,
\href{http://dx.doi.org/10.1088/0264-9381/15/6/005}{Class.\ Quantum
Grav. {\bf 15} (1998) 1463}

\bibitem{Y-M-Energy}
J.\ Yang and Y.\ Ma, Quasi-local energy in loop quantum gravity,
\href{http://dx.doi.org/10.1103/PhysRevD.80.084027}{Phys.\ Rev.\ D {\bf 80} (2009)
084027}

\bibitem{Mar-Energy}
S.\ Major, Quasilocal energy for spin-net gravity,
\href{http://dx.doi.org/10.1088/0264-9381/17/6/311}{Class.\ Quantum
Grav. {\bf 17} (2000) 1467}

\bibitem{AshVar}
A.\ Ashtekar, New Hamiltonian formulation of general relativity,
\href{http://dx.doi.org/10.1103/PhysRevD.36.1587}{Phys.\ Rev.\ D
{\bf 36} (1987) 1587}

\bibitem{Barbero}
J.~F.\ Barbero~G., Real Ashtekar variables for Lorentzian signature
space-times,
\href{http://dx.doi.org/10.1103/PhysRevD.51.5507}{Phys.\ Rev.\ D
{\bf 51} (1995) 5507}

\bibitem{AI}
A.\ Ashtekar and C.~J.\ Isham, Representations of the holonomy
algebras of gravity and non-abelean gauge theories,
\href{http://dx.doi.org/10.1088/0264-9381/9/6/004}{Class.\ Quantum
Grav. {\bf 9} (1992) 1433}

\bibitem{AL}
A.\ Ashtekar and J.\ Lewandowski, Projective techniques and
functional integration for guage theories,
\href{http://dx.doi.org/10.1063/1.531037}{J.\ Math.\ Phys. {\bf 36}
(1995) 2170}

\bibitem{Uniq}
J.\ Lewandowski, A.\ Okolow, H.\ Sahlmann and T.\ Thiemann,
Uniqueness of diffeomorphism invariant states on holonomy-flux
algebras,
\href{http://dx.doi.org/10.1007/s00220-006-0100-7}{Commun.\ Math.\
Phys. {\bf 267} (2006) 703}

\bibitem{Major-Direc}
S. Major, Operators for quantized directions,
\href{http://dx.doi.org/10.1088/0264-9381/16/12/307}{Class. Quant.
Grav. {\bf 16} (1999) 3859}

\bibitem{Brunn}
J.\ Brunnemann and D.\ Rideout, Properties of the volume operator in
loop quantum gravity I: Results,
\href{http://dx.doi.org/10.1088/0264-9381/25/6/065001}{Class.\
Quant.\ Grav. {\bf 25}
(2008) 065001}\\
J.\ Brunnemann and D.\ Rideout, Properties of the volume operator in
loop quantum gravity II: Detailed presentation ,
\href{http://dx.doi.org/10.1088/0264-9381/25/6/065002}{Class.\
Quant.\ Grav. {\bf 25} (2008) 065002}


\bibitem{G-T-check}K.\ Giesel and T.\ Thiemann, Consistency check on volume and triad
operator quantization in loop quantum gravity: I,
\href{http://dx.doi.org/10.1088/0264-9381/23/18/011}{Class. Quant.
Grav. {\bf 23} (2006) 5667}\\
K.\ Giesel and T.\ Thiemann, Consistency check on volume and triad
operator quantization in loop quantum gravity: II,
\href{http://dx.doi.org/10.1088/0264-9381/23/18/012}{Class. Quant.
Grav. {\bf 23} (2006) 5693}

\bibitem{Tik}A.\ Tikhonov, On the stability of inverse problems, Dokl. Akad.
Nauk SSSR {\bf 39} (1943) 195

\bibitem{F-semiVol}C.\ Flori and T.\ Thiemann, Semiclassical analysis of the loop quantum gravity volume operator: I. Flux coherent
states, \href{http://arxiv.org/abs/0812.1537}{arXiv:0812.1537}\\
C.\ Flori, Semiclassical analysis of the loop quantum gravity volume
operator: Area coherent states,
\href{http://arxiv.org/abs/0904.1303}{arXiv:0904.1303}
\end{thebibliography}
\end{document}